\documentclass[aps,pra,twocolumn,showpacs,superscriptaddress,a4paper,groupedaddress]{revtex4-1}
\usepackage{graphicx}    
\usepackage{dcolumn}    
\usepackage{bm}             
\usepackage{amssymb}   
\usepackage{amsmath}    
\usepackage{subfigure}    
\usepackage{braket}
\usepackage{mathrsfs}
\usepackage[export]{adjustbox}
\usepackage[toc,page]{appendix}
\begin{document}

\title{Collective Dipole-Dipole Interactions in an Atomic Array}
\author{R.T. Sutherland$^{1}$}
\email{rsutherl@purdue.edu}

\author{F. Robicheaux $^{1,2}$}
\email{robichf@purdue.edu}
\affiliation{$^{1}$Department of Physics and Astronomy, Purdue University, West Lafayette IN, 47907 USA}
\affiliation{$^{2}$Purdue Quantum Center, Purdue University, West Lafayette,
Indiana 47907, USA}

\date{\today}

\begin{abstract}
The coherent dipole-dipole interactions of atoms in an atomic array are studied. It is found that the excitation probability of an atom in an array parallel to the direction of laser propagation ($\boldsymbol{\hat{k}}$) will either grow or decay logarithmically along $\boldsymbol{\hat{k}}$, depending on the detuning of the laser. The symmetry of the system for atomic separations of $\delta r = j\lambda/2$, where $j$ is an integer, causes the excitation distribution and scattered radiation to abruptly become symmetric about the center of the array. For atomic separations of $\delta r < \lambda/2$, the appearance of a collection of extremely subradiant states ($\Gamma\sim 0$), disrupts the described trend. In order to interpret the results from a finite array of atoms, a band structure calculation in the $N\rightarrow \infty$ limit is conducted where the decay rates and the collective Lamb shifts of the eigenmodes along the Brillouin zone are shown. Finally, the band structure of an array strongly affects its scattered radiation, allowing one to manipulate the collective Lamb shift as well as the decay rate (from superradiant to subradiant) by changing the angle of the driving laser. 
\end{abstract}
\maketitle

\section{Introduction}\label{intro}
The coherent nature of the coupling between a collection of radiators and the electromagnetic field \cite{dicke1954} has proven to be a fruitful field of study for over 60 years. Despite its history, the study of collective radiation is still providing important insights into quantum optics, leading to deeper understandings of systems such as cold atom clouds \cite{sutherland2016, bromley2016, de2014, javanainen2014, bienaime2012, ido2005, roof2016, kaiser2016}, waveguides \cite{goban2015}, quantum information \cite{yavuz2015}, and biophysics \cite{monshouwer1997, palacios2002}. Recently, the interplay between the collective Lamb shift \cite{meir2014, scully2009, rohlsberger2010}, the energy shift due to the exchange of virtual photons between radiators \cite{gross1982}, and its relationship to superradiance/subradiance \cite{scully2007, kaiser2015, scully2015, wang2015} has produced a plethora of new physics. In particular, the study of the large-scale coherent build-up of \textit{forward} photon emission in cold atom clouds has shown that coherent dipole-dipole interactions can cause dramatic changes to the nature of the cloud's scattered radiation \cite{svidzinsky2008, rouabah2014, sutherland2016, bromley2016, scully2006}. Recently, it has also been noted that coherent dipole-dipole interactions can cause the excitation distribution of a cigar-shaped cloud to deviate from the results predicted by the Beer-Lambert law \cite{sutherland2016}.
\begin{figure}[h!]
	\includegraphics[width=0.35\textwidth]{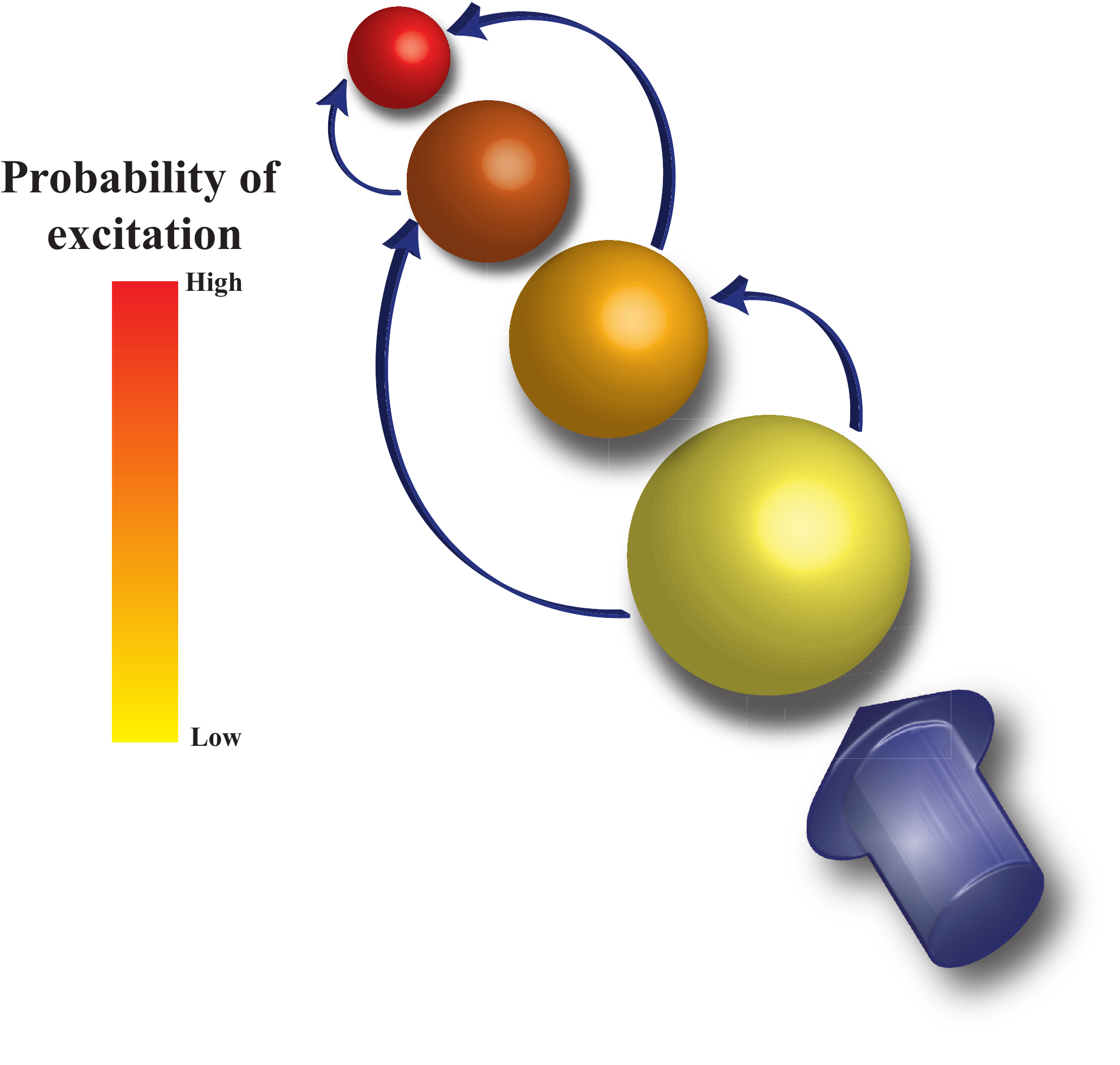}
	\caption{When a red detuned laser drives an array of atoms, the probability of excitation grows logarithmically down the line of the array. This effect and the others discussed in this paper, are due to the spatially dependent phase correlations between the driving laser and the dipole radiation.} 
	\label{fig:pretty}
\end{figure}

Collective interactions in atomic arrays are also well-studied and have been shown to produce exciting effects, such as the appearance of superradiant and subradiant eigenmodes \cite{porras2008, mewton2007, clemens2003, ostermann2012, zoubi2012}. However, these studies do not fully address the physics of position dependent phase correlations in the array. When an entire system is illuminated by a laser, it excites the atoms to a state where the phase of the excitation probability amplitude of each atom is proportional to the laser's own phase (a timed-Dicke state) \cite{scully2006}. This produces coherences that dramatically change the photon scattering \cite{svidzinsky2008,scully2009,scully2015,svidzinsky2010}. This effect has recently been studied by considering the emission of timed-Dicke states when an array has spacings ($\delta r$) much less than the resonance wavelength ($\lambda$), where only nearest neighbor interactions are considered \cite{liao2014}. However, the limit $\delta r \ll \lambda$ is not relevant to most experimental setups. Also, in this regime the near field term in the dipole field propagator (see Eq.~(\ref{eq:g})) overshadows the physics resulting from the coherent buildup of the $\propto 1/r$ term over the entire array. When $\delta r \sim \lambda$ or larger, even though the individual dipole-dipole interactions are small, they can coherently build over a large sample and cause non-negligible effects. For example, this interaction has been shown to produce an observable shift (collective Lamb shift) in the light scattered from an array of ions when $\delta r > \lambda$ \cite{meir2014}. In this paper, we show that when both the phase correlations caused by the driving laser and the collective dipole-dipole interactions between \textit{all} atoms in an array are considered, they produce novel physics that allows for the manipulation of the atoms' scattered light and excitation probability.

Specifically, it is shown that an array's probability distribution, when driven by a laser parallel to the array, is highly dependent on the detuning of the laser. For red (blue) detuned light, the dipole radiation in the direction of the laser causes the probability of excitation to increase (decrease) logarithmically  (see Fig. \ref{fig:pretty}). For atomic separations ($\delta r$) of $j\lambda/2$, where $j$ is an integer, the logarithmic excitation distribution along the direction of the laser $(\boldsymbol{\hat{k}})$ rapidly changes so that it becomes completely symmetric about the center of the array.
The results described do not hold for ($\delta r < \lambda /2$) due to the existence of a collection of non-radiating ($\Gamma \sim 0$) eigenmodes in this regime. In the interest of understanding the eigenmodes and eigenvalues of a finite array of atoms, the $N \rightarrow \infty$ limit is then implemented in order to conduct a band structure calculation. This calculation gives both the collective Lamb shift and the decay rate of the eigenmodes along the Brillouin zone. It is then shown that the appearance and disappearance of Bragg diffraction peaks causes the eigenmodes to discontinuously jump from subradiant (superradiant) to superradiant (subradiant), depending on $\delta r$, when plotted along the Brillioun zone. Lastly, it is demonstrated how the band structure of a long array of atoms strongly affects the light scattered from the array, allowing one to probe subradiant eigenmodes in a straightforward manner.

\section{Theory/Methods} \label{theory}
For a weak laser, a collection of two-level atoms polarized in the $\hat{\boldsymbol{x}}$ direction can be treated as coupled damped harmonic oscillators \cite{javanainen2014, jenkins2012, svidzinsky2010, ruostekoski1997}, 
\begin{eqnarray}
\label{eq:a}
\dot{a}_{j}(t)&=\nonumber&(i\Delta - \Gamma /2)a_{j}(t) -  i(d /\hbar)E(\boldsymbol{r}_{j}) \\ &-& (\Gamma /2)\sum_{m\neq j}G({\boldsymbol{r}}_{m} - {\boldsymbol{r}}_{j})a_{m}(t),
\end{eqnarray}
where $a_{j}$ represents the polarization amplitude of the $j^{th}$ atom, d is the electric dipole matrix element, $E(\boldsymbol{r}_{j}) = E_{0}e^{i\boldsymbol{k}\cdot \boldsymbol{r}_{j}}$ is the the laser field at atom $j$, $\Delta$ is the detuning, $\Gamma$ is the single atom decay rate, and $G(\boldsymbol{r})$ is the usual dipole field propagator \cite{jackson1999},
\begin{equation}
G(\boldsymbol{r}) = \frac{3e^{ikr}}{2ikr} \{[1 - (\hat{\boldsymbol{r}}\cdot\hat{\boldsymbol{x}})^{2}] + [1 - 3(\hat{\boldsymbol{r}}\cdot\hat{\boldsymbol{x}})^{2}][\frac{i}{kr} - \frac{1}{(kr)^{2}}]\},
\label{eq:g}
\end{equation}
where $r = |\boldsymbol{r}|$, and $\hat{\boldsymbol{r}}$ is the vector $\boldsymbol{\hat{r}}$ = $\boldsymbol{r}$/r. These coupled equations can be rewritten in matrix-vector form:
\begin{equation}\label{eq:matrix_eq}
\dot{\boldsymbol{a}}=\underline{\boldsymbol{M}}\boldsymbol{a} - i\frac{d}{\hbar}\boldsymbol{E}
\end{equation}
and the steady state solution ($\dot{\boldsymbol{a}}=0$) may be obtained by inverting a symmetric $N\times N$ matrix.

For the band structure calculations of Section \ref{sec:diffraction}, the $N \rightarrow \infty$ limit is implemented in order to calculate the complex eigenvalues of $\underline{\boldsymbol{M}}$. In this limit, the translational symmetry of the system may be used in order to rewrite the eigenvalue problem:
\begin{equation}
\sum_{n} M_{mn}a_{n} = \epsilon_{q} a_{m}
\end{equation}
as
\begin{eqnarray}\label{eq:band_eq}
\sum_{n} M_{mn}e^{in\delta r q}a_{0} &=&\epsilon_{q} e^{im\delta r q}a_{0} \\
\sum_{n} M_{mn}e^{i(n-m)\delta r q} &=&\epsilon_{q},
\end{eqnarray}
so that the calculation reduces to an infinite sum that converges for most values of $q$ (see Section \ref{sec:diffraction}). Here the decay rate and the collective Lamb shift of the $q^{th}$ eigenmode can be obtained from $\epsilon_{q}$'s real and imaginary parts respectively.

\section{Excitation Distribution} \label{line}
\subsection{Numerical Results}\label{sec:numerical}

\begin{figure}[h]
	\includegraphics[width=0.45\textwidth]{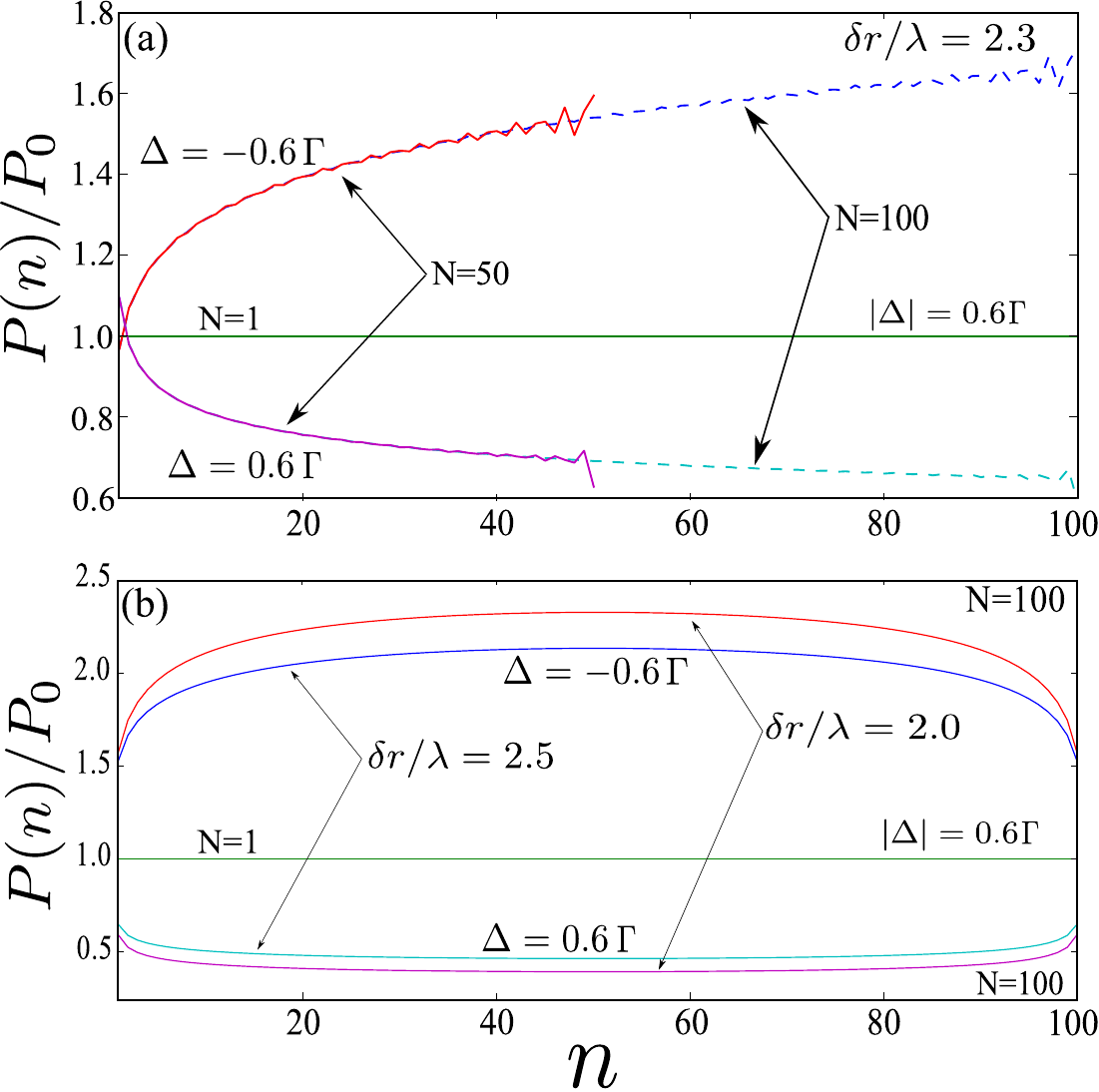}
	\caption{(a) Probability of excitation of atom $n$ ($P(n)$) divided by the single atom excitation probability ($P_{0}$) for that detuning ($\Delta$). For the top two plots, $P(n)/P_{0}$ is shown for a detuning of $-0.6\Gamma$ for 50 and 100 atoms. Note that except for small oscillations, both plots lie on top of each other for up to $n$=50. This is due to the highly forward character of the coherent interactions. For the bottom two plots, $P(N)/P_{0}$ is shown for a detuning of $+0.6\Gamma$ for 50 and 100 atoms. The non-interacting probability for the same rabi frequency and $|\Delta|$ is shown for reference. (b) The symmetric probability distribution present for $\delta r \rightarrow \frac{j\lambda}{2}$ is shown for red and blue detunings as well as integer and half integer values of $\delta r$. Note that the difference in the the integer and half integer plot is mainly due to the differences in $\delta r$.} 
	\label{fig:perfect_line}
\end{figure}

An understanding of the highly directional nature of the interactions that add coherently can be gained by noting the similarities in the phases accumulated between the driving laser and the dipole-dipole interactions \cite{sutherland2016}. Essentially, the phase a laser will accumulate when going from atom $i$ to atom $j$ is $e^{i\boldsymbol{k}\cdot(\boldsymbol{r}_{i} - \boldsymbol{r_{j}})}$, while the phase accumulated in a photon exchanged by the two atoms is $e^{ik|\boldsymbol{r}_{i} - \boldsymbol{r_{j}}|}$. These two phases are equivalent when $(\boldsymbol{r}_{i}-\boldsymbol{r}_{j})$ is parallel to $\hat{k}$. Because of this, all of the radiation and virtual photon exchanges along $\hat{\boldsymbol{k}}$ add coherently, while they add incoherently along $-\hat{\boldsymbol{k}}$ relative to atom $n$. Therefore, the excitation probability of atom $n$, ($P(n)$), depends mainly on the dipole-dipole interactions from the $n-1$ atoms in the $-\boldsymbol{\hat{k}}$ direction relative to $n$. This is shown in Fig. \ref{fig:perfect_line}(a), where except for small oscillations caused by reflections off of the end of the array, the value of $P(n)$ follows approximately the same curve for an array of 50 atoms as an array of 100 atoms for both red and blue detunings. Note that this mechanism of coherence in the forward direction is only true when $\delta r \ne j\lambda/2$, as will be discussed shortly.

The nature of $P(n)$ can be intuitively understood by considering that the $n^{th}$ atom in the array will only see a significant contribution of electric field from the driving laser and the $n-1$ atoms located in the $-\hat{\boldsymbol{k}}$ direction relative to it, as well as the fact that the dominant term in the dipole-dipole interaction is $\propto 1/(kr)$ for the values of $\delta r$ described here. It can now be seen that atom $n$ will feel a sum of dipole-dipole interactions, that add either constructively or destructively with the driving laser, with a magnitude:
\begin{equation}\label{eq:approx}
P(n) \propto \frac{1}{k\delta r}\sum_{m < n} \frac{1}{m},
\end{equation}
which is $\sim\ln (n)$ for large values of $n$. This is approximately the form of $P(n)$ shown in Fig. \ref{fig:perfect_line}(a). 

The above description holds only for values of $\delta r \neq \frac{j \lambda}{2}$, where $j$ is an integer. However, the only terms in Eq.~(\ref{eq:a}) distinguishing $a_{1}$ from $a_{N}$, $a_{2}$ from $a_{N-1}$, etc... are the phase factors $e^{ikj \delta r}$. If $e^{ikj \delta r} \rightarrow \pm 1$ Eq.~(\ref{eq:a}) is symmetric about the center of the array. Resultantly when $\delta r \rightarrow \frac{j\lambda}{2}$, $P(n)$ becomes completely symmetric about the center of the array. Because of this symmetry, the only parameter determining the value of $P(n)$ is the total magnitude of all the dipole-dipole interactions atom $n$ experiences. For red detunings, it has already been established that all of the dipole-dipole interactions add in phase with the laser. As seen in Fig. \ref{fig:perfect_line}(b), this causes the atoms experiencing the strongest interactions (atoms in the center of the array)  to be the most excited. For blue detunings dipole-dipole interactions add destructively, here Fig. \ref{fig:perfect_line}(b) shows that the atoms in the center are the least excited. Because both the forward and backward dipole-dipole interactions now add coherently, the dependence described in Eq.~(\ref{eq:approx}) becomes:
\begin{equation}\label{eq:other_approx}
P(n) \propto \frac{1}{k\delta r}\sum_{m \neq n} \frac{1}{|m-n|}.
\end{equation}

\begin{figure}[h!]
	\includegraphics[width=0.45\textwidth]{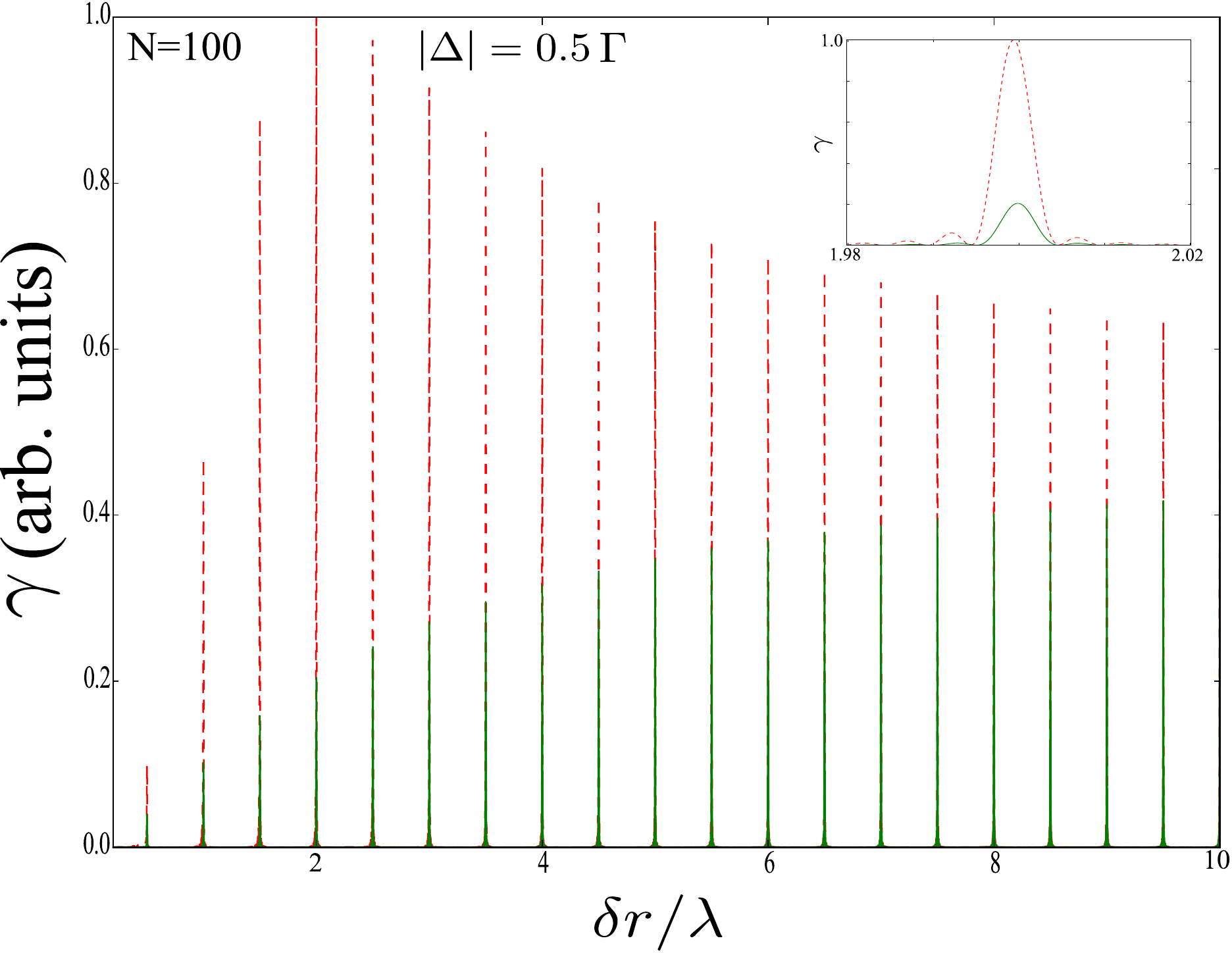}
	\caption{Photon scattering rate ($\gamma$) in the $-\boldsymbol{\hat{k}}$ direction versus array spacing ($\delta r$) for an array of 100 atoms for values of detuning $\Delta = 0.5\Gamma$ (green solid line) and $\Delta = -0.5\Gamma$ (red dotted line). The inset shows the same graph zoomed in around $\delta r = 2\lambda$. Note that the difference in the heights of the peaks is mainly due to the fact that the collective Lamb shift of the driven eigenmode changes with $\delta r$.} 
	\label{fig:cbs}
\end{figure}

The symmetry about the center of the array also causes a large increase in the coherent backscattering (see Fig. \ref{fig:cbs}). Normally the phase correlations of an array of atoms parallel to $\boldsymbol{\hat{k}}$ only allow for Coherent Forward Scattering \cite{rouabah2014,scully2009}, however because of this symmetry about $\pm \boldsymbol{\hat{k}}$ when $\delta r \rightarrow \frac{j\lambda}{2}$ light scattered in the $-\boldsymbol{\hat{k}}$ direction also adds coherently, causing a diffraction peak. The inset in Fig. \ref{fig:cbs} shows a close up of one of the diffraction peaks. Near a given peak, the coherent backscattering has the approximate form: $\{j_{0}(kN\alpha)\}^{2}$, where $\alpha = \delta r - j\lambda/2$ and $j_{0}(x)$ is the zeroth spherical Bessel function. This can be shown using Eq.~(\ref{eq:direction1}) and making the approximation that the atoms are in the timed-Dicke state ($a_{j} = |a_{0}|e^{ik j\delta r}$) caused by a laser propagating in the $+\hat{\boldsymbol{k}}$ direction. 

The differences in the heights of the diffraction peaks shown in Fig. \ref{fig:cbs}, are mainly caused by the fact that the collective Lamb shift of the eigenmode that the laser drives changes with $\delta r$. In Fig. \ref{fig:cbs}, the intensity of successive diffraction peaks increases until $\delta r = 2\lambda$, where it then begins to decrease. For different laser detunings, this pattern would follow a different form.  A similar effect can be seen in in Fig. \ref{fig:perfect_line} where the magnitude of the the change in $P(n)$ for a given array is larger for red detunings than blue detunings of the same magnitude, since the resonance frequency for an array of atoms parallel to $\boldsymbol{\hat{k}}$ is red shifted (see Sec. \ref{sec:diffraction}).

Figure \ref{fig:perfect_line}(a) is for $\delta r=2.3\lambda$ and detunings of $|\Delta| = 0.6$, which for red detunings shows a $\sim 70 \%$ difference between the first and last atoms. However, due to the long-range nature of the coherent build-up of dipole radiation, the logarithmic growth of $P(n)$ does not saturate for large values of $\delta r$. For example, in Fig. \ref{fig:analitic}(b), a $\sim 2.6 \%$ growth in $P(n)$ is seen for $\delta r = 50.3\lambda$. In fact, until other timescales, such as retardation effects, become important there is no value of $\delta r$ where this logarithmic growth doesn't, in principle, happen. 

The nature of $P(n)$ described here is qualitatively valid in non-ideal circumstances. This is tested using Monte Carlo routines where the filling factor and the randomness of the position of each atom is varied. The magnitude of the overall growth for non-integer wavelengths is approximately proportional to the filling factor of the sample. For example, if an experiment would have produced an array where the first and last atoms have an excitation probability that differs by 50\%, a filling factor of 0.5 causes the overall effect to reduce to $\sim 25\%$. It is also found that for $\delta r \neq \frac{j\lambda}{2}$ when each atom's $x,y,z$ values are allowed to randomly vary, the noise of a given array's $P(n)$ increases while the average value of $P(n)$ does not change until the randomness of the atoms' positions are allowed to vary distances comparable to $\delta r$, not $\lambda$. For example if $\delta r = 20.3$, the sample is significantly more resilient to random positions than if $\delta r = 2.3$. However, it is found that the described symmetry for $\delta r = \frac{j\lambda}{2}$ is more sensitive to non-ideal scenarios. Unlike the logarithmic buildup, the robustness of the symmetry about the center of the array does not seem to depend on the value of $\delta r$. It is found that for all spacings, letting the atom positions randomly vary more than $\sim 0.3\lambda$, causes the symmetric $P(n)$ distribution to begin to approach the logarithmic function seen for $\delta r \neq \frac{j\lambda}{2}$. Note that the probability distributions described here only occur when $\boldsymbol{\hat{k}}$ is parallel to the array.

\subsection{Analytic Derivation of Excitation Distribution}\label{sec:analitic}
\begin{figure}[h!]
	\includegraphics[width=0.45\textwidth]{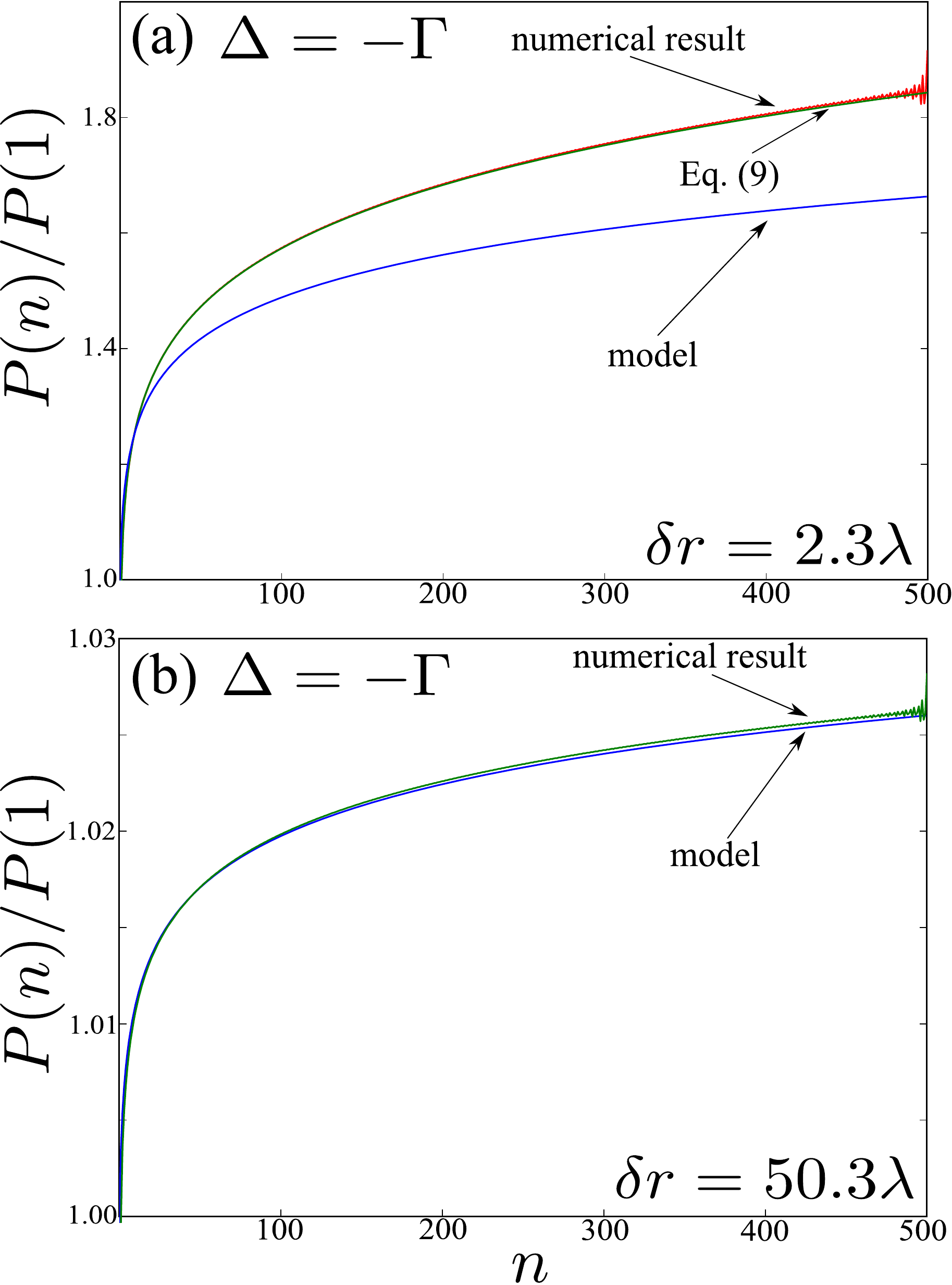}
	\caption{The probability of excitation of atom $n$ ($P(n)$) divided by the probability of excitation of the first atom ($P(1)$), for an array containing 500 atoms. (a) Comparison between the full numerical calculation (red), Eq.~(\ref{eq:system}) (green), and our analytic model (blue) for $\delta r = 2.3\lambda$. Since the approximations made in the analytic derivation hold only for $k\delta r \gg 1$, the model is only qualitatively accurate in this regime. Note that when $P(n)$ is generated from Eq.~(\ref{eq:system}), it lies on top of the full calculation (excluding small oscillations at large $n$) for all $\delta r > \lambda$. (b) Comparison between the full numerical calculation (green) and the analytic model (blue) for $k\delta r \gg 1$. This illustrates the fact that the analytic model approaches the quantitive numerical result for this condition, as well as the fact that the logarithmic growth holds for very large array spacings. Note that these results are only true for $\delta r \ne \frac{j\lambda}{2}$.} 
	\label{fig:analitic}
\end{figure}

In this section, the approximation that dipole-dipole interactions adding incoherently (occuring in the $-\hat{\boldsymbol{k}}$ direction) are negligible, is implemented in order to derive an equation for $P(n)$ analytically. Note that this is only valid when $\delta r \neq \frac{j\lambda}{2}$. Neglecting all incoherent interactions allows us to replace $\boldsymbol{\underline{M}}$ in Eq.~(\ref{eq:matrix_eq}) with a lower triangular matrix. When the $\dot{\boldsymbol{a}} \rightarrow 0$ limit is taken, solving for $\boldsymbol{a}$ is reduced to solving the system of equations:
\begin{eqnarray}\label{eq:system}
0 &=& (i\Delta - \frac{\Gamma}{2})a_{1} - i(d /\hbar)E(\boldsymbol{r}_{1}) \nonumber \\
0 &=& (i\Delta - \frac{\Gamma}{2})a_{2} - i(d /\hbar)E(\boldsymbol{r}_{2})  - \frac{\Gamma}{2}G(\delta r)a_{1} \nonumber\\
0 &=& (i\Delta - \frac{\Gamma}{2})a_{3} - i(d /\hbar)E(\boldsymbol{r}_{3}) - \frac{\Gamma}{2}G(\delta r)a_{2}
- \frac{\Gamma}{2}G(2\delta r)a_{1} \nonumber \\ 
&...&,
\end{eqnarray}
where $a_{n}$ is the $n^{th}$ component of $\boldsymbol{a}$. This can be solved for $a_{1}$, which can be plugged into the equation for $a_{2}$ etc... The approximation indicated in Eq.~(\ref{eq:system}) remains quantitatively accurate when $\delta r > \lambda$. This can be seen in Fig. \ref{fig:analitic}(a) where, except for small oscillations due to reflections off the end of the array, the curve produced by solving Eq.~(\ref{eq:system}) remains nearly identical to the full numerical result. Assuming Eq.~(\ref{eq:system}), and keeping only the first order terms of $1/(k\delta r)$ allows one to obtain a closed form solution for $a_{n}$:
\begin{equation}\label{eq:solved_system2}
a_{n}e^{-ik(n-1) \delta r} =\frac{idE_{0}/\hbar}{i\Delta - \Gamma/2}\Big(1 - \frac{3i\Gamma}{4k\delta r(i\Delta - \Gamma/2)} \sum_{m=1}^{n-1}\frac{1}{m}\Big).
\end{equation}
This equation holds in the limit $k\delta r \gg 1$, and causes $P(n)$ to grow for red detunings and diminish for blue detunings. As seen in Fig. \ref{fig:analitic}, for smaller spacings the simple model used to derive Eq.~(\ref{eq:solved_system2}) becomes only qualitative. This is because in this regime, higher order $1/(k\delta r)$ terms matter. As $\delta r$ becomes smaller than $\lambda$, individual dipole-dipole interactions grow and despite the fact that they add incoherently, the contributions of a couple of large interactions in the $-\hat{\boldsymbol{k}}$ direction begin to become significant, causing Eq.~(\ref{eq:system}) to break down. 

\section{$\delta r < \lambda/2$ Behavior} \label{sec:short}
\begin{figure}[h!]
	\includegraphics[width=0.45\textwidth]{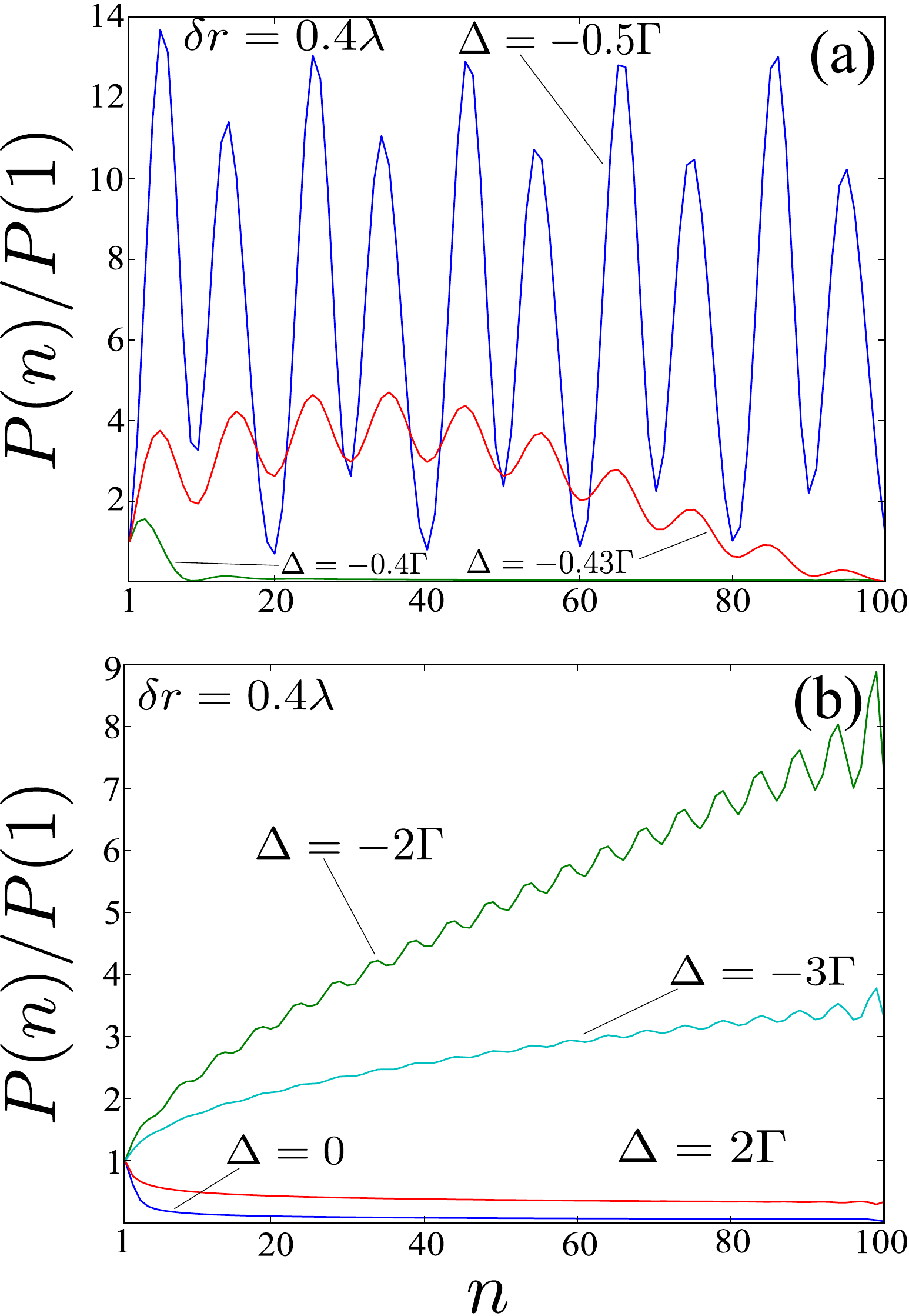}
	\caption{The probability of excitation of atom $n$ ($P(n)$) divided by the probability of excitation of atom $1$ ($P(1)$). (a) Excitation distribution for a laser resonant with the collection of subradiant states ($Re(\epsilon_{n}) \sim 0$) for an array of 100 atoms separated by a distance of $0.4\lambda$. Over the range of laser detunings $\Delta = -0.4\Gamma- \ -0.9\Gamma$ the behavior is drastically altered due to the almost complete lack of decay by photon emission. (b) The same array of atoms for laser detunings that do not lie within ($\Delta \sim -0.4\Gamma- \ -0.9\Gamma$) show qualitatively similar results to those described previously.}
	\label{fig:short_sep}
\end{figure}

\begin{figure}[h!]
	\includegraphics[width=0.45\textwidth]{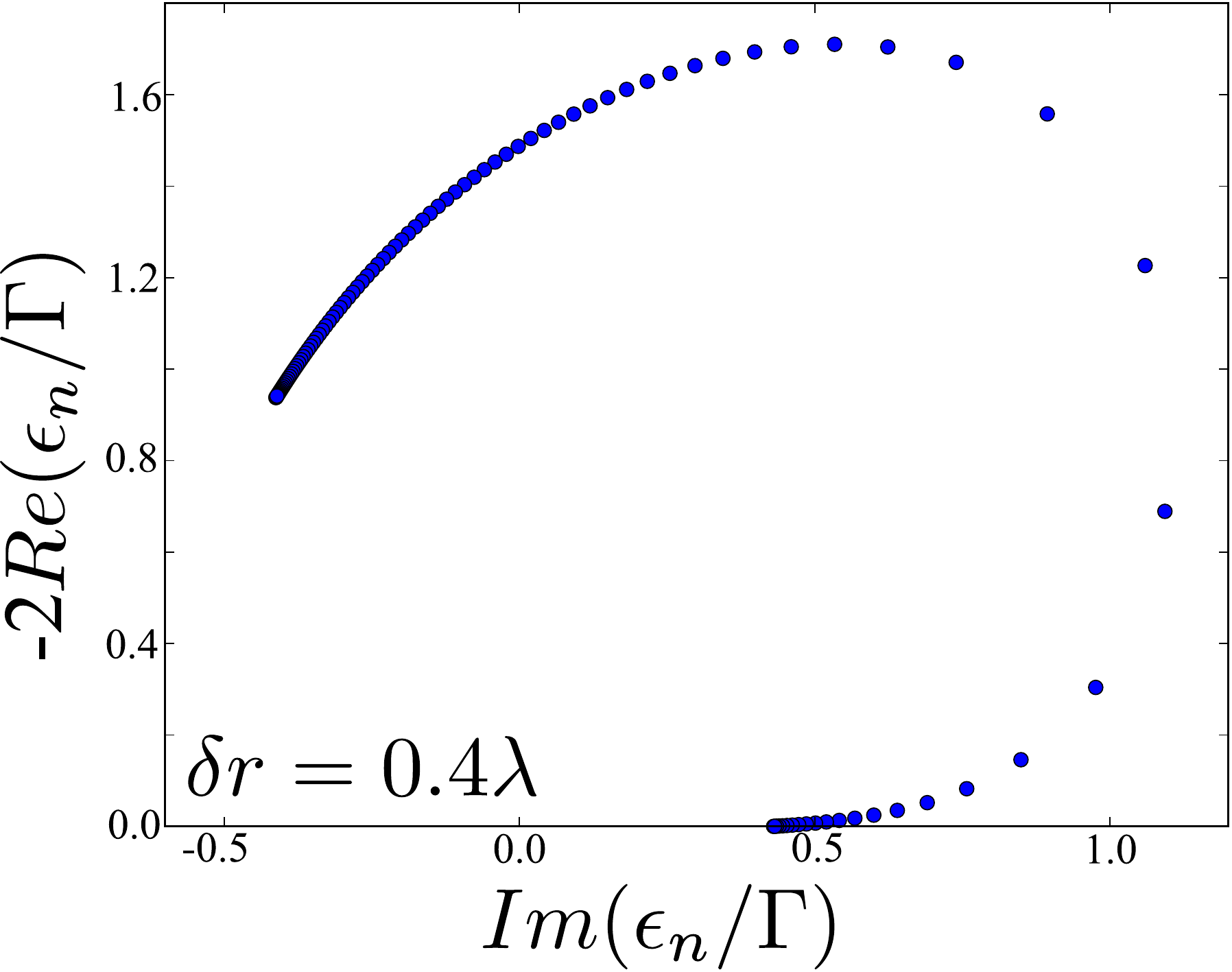}
	\caption{The decay rate ($-2Re(\epsilon_{n})$) and the collective Lamb shift ($Im(\epsilon_{n})$) of the eigenmodes of an array  divided by the single atom decay rate ($\Gamma$), where $\delta r = 0.4\lambda$, $\Delta = 0$ and N = 100. Note the collection of eigenvalues near the $-2Re(\epsilon_{n}) = 0$ axis.}
	\label{fig:eigen}
\end{figure}

The excitation distribution described above breaks down for values of $\delta r < \lambda/2$. This may be understood by considering the distribution of the eigenvalues ($\epsilon_{n}$) of $\boldsymbol{\underline{M}}$. Since $\boldsymbol{\underline{M}}$ is complex symmetric rather than Hermitian, the values of $\epsilon_{n}$ are complex. Here, $-2Re(\epsilon_{n})$ is the $n^{th}$ eigenmode's decay rate, while $Im(\epsilon_{n})$ is its collective Lamb shift. When $\delta r < \lambda/2$, there exists a collection of non-radiating ($-2Re(\epsilon_{n})\sim 0$) eigenmodes of $\boldsymbol{\underline{M}}$ within a relatively small energy range (see Fig. \ref{fig:eigen}). This effect is well-known, and has been used in systems such as arrays of metallic nanospheres \cite{burin2004, chui2015, halir2015}, where it has been shown that these eigenmodes may be used for their optical transport properties. The same physics strongly affects the excitation distribution of an array driven by a laser. In Fig. \ref{fig:short_sep}, the excitation distribution for an array of 100 atoms, where $\delta r = 0.4\lambda$, is shown for various values of $\Delta$. In Fig. \ref{fig:short_sep}(a), it may be seen that the values of $\Delta$ that lie within the energy range of the collection of subradiant states ($\Delta = -0.4\Gamma- \ -0.9\Gamma$) produce very different excitation distributions than the logarithmic ones described in the previous sections. However, Fig. \ref{fig:short_sep}(b) shows that once the laser is off resonance with the collection of subradiant states, the distribution becomes logarithmic again.

Figure \ref{fig:eigen} shows the decay rates ($-2Re(\epsilon_{n})$) and collective Lamb shifts ($Im(\Delta_{n})$) for the eigenmodes of an array where  $\delta r = 0.4\lambda$, $\Delta = 0$, and $N=100$. Here it can be seen that the collection of eigenmodes whose eigenvalues ($\epsilon_{n}$) occur near the $-2Re(\epsilon_{n}) = 0$  axis are within a very narrow range of energies. It should be noted that for Fig. \ref{fig:eigen}, a positive value of $Im(\epsilon_{n})$ corresponds to an eigenvalue that would be on resonance with a red detuned laser. This is because the value of $\Delta$ is present on all of the diagonals of $\underline{\boldsymbol{M}}$ and would resultantly shift all of the imaginary parts of the eigenvalues towards $Im(\epsilon_{n}) = 0$, i.e. resonance. Even though populating an individual subradiant state can be difficult due to the narrowness in energy of its photon scattering cross section, the fact that they all occur in a very small energy range causes them to be the dominant feature of the steady state solution for values of $\Delta$ within their energy band. The presence of $-2Re(\epsilon_{n}) \sim 0$ decay modes will be explained in Section \ref{sec:diffraction}.

\section{Band Structure}\label{sec:diffraction}
In order to understand the eigenmodes and eigenvalues of this system, we examine its band structure in the $N \rightarrow \infty$ limit. Since in Eq.~(\ref{eq:a}) there is only one oscillator per atom, there is only one band of eigenvalues. For a given eigenvalue ($\epsilon_{q}$), again the real part corresponds to the decay rate of the $q^{th}$ eigenmode \textit{divided by two} ($-2Re(\epsilon_{q}) = \Gamma_{q}$), while the imaginary part corresponds to the collective Lamb shift of the $q^{th}$ eigenmode ($\Delta_{q}$). Since $i\Delta$ occurs in every diagonal element of $\underline{\boldsymbol{M}}$, its value will only produce a uniform shift to the imaginary part of every $\epsilon_{q}$, thus it is taken to be zero. Note that values of $\Delta_{q}$ that are positive correspond to redshifts in the resonance line of that eigenmode.

\begin{figure}[h!]
	\includegraphics[width=0.45\textwidth]{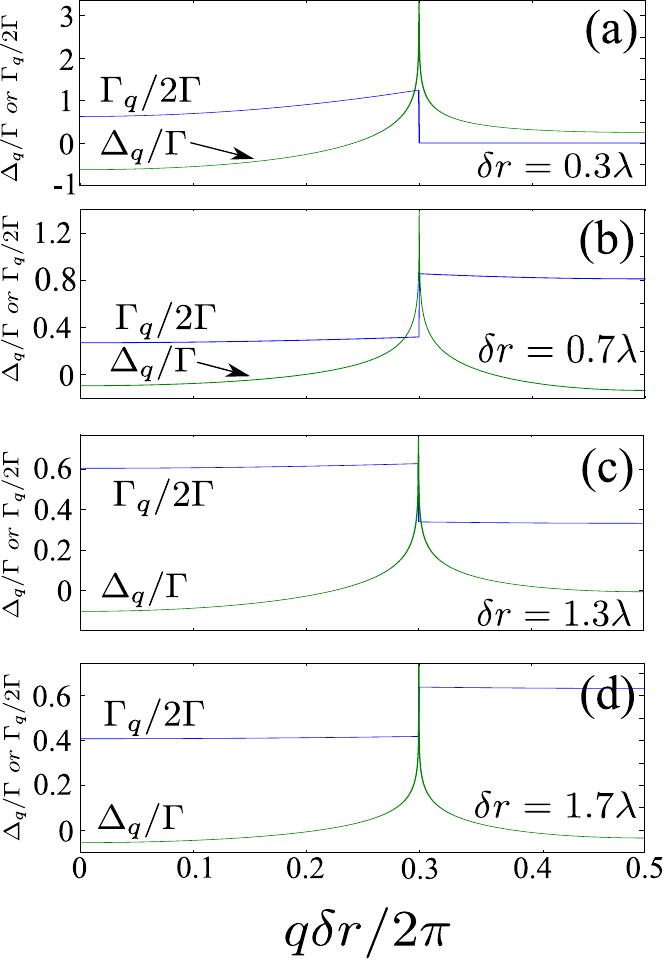}
	\caption{The values of $\Gamma_{q}$ and of $\Delta_{q}$ (the collective Lamb shift), are plotted versus values of $0 \leq q\delta r / 2\pi < 0.5$. (a) Shows this for array spacings of $\delta r = 0.3\lambda$, (b) $\delta r = 0.7\lambda$, (c) $\delta r = 1.3\lambda$, and (d) $\delta r = 1.7\lambda$. Note that for all four graphs the collective Lamb shift diverges at $q\delta r/2\pi = 0.3$, while $\Gamma_{q}$ gives a discontinuous jump, caused by the appearance or disappearance of a Bragg peaks.}
	\label{fig:band}
\end{figure}

In Fig. \ref{fig:band}, the values of $\Gamma_{q}$ and $\Delta_{q}$ are plotted for the positive half of the Brillouin zone ($0 \geq q < \pi / \delta r$). The negative values of $q$ may be omitted since the calculation is symmetric and yields the same results. In Fig. \ref{fig:band}(a), a discontinuity is seen at $k\delta r = q\delta r$, when the value of $\Gamma_{q}$ drops to 0, changing the eigenmode from superradiant to non-radiant. At the same time in Fig. \ref{fig:band}(b), when $k\delta r + (q\delta r) = 2\pi$ the value of $\Gamma_{q}$ shows a discontinuity from a subradiant mode to a superradiant mode. The same pattern occurs for larger array spacings. In Fig. \ref{fig:band}(c), when $(k\delta r) - 2\pi = q\delta r$ the eigenmodes shift from superradiant to subradiant, while in Fig. \ref{fig:band}(d), when $k\delta r + (q\delta r) = 4\pi$ the eigenmodes jump from subradiant to superradiant. While the magnitude of each discontinuous jump of $\Gamma_{q}$ decreases with $\delta r$, approaching $\Gamma_{q}=\Gamma$ when $\delta r \rightarrow \infty$, the described band structure pattern repeats for every integer increase in $\delta r/\lambda$. This can be understood in terms of the appearance and disappearance of Bragg diffraction peaks. This phenomena was somewhat addressed in \cite{porras2008}, but the effect of different phase correlations (values of $q$) on the decay rate was not.

Figure \ref{fig:band} can be understood by examining the photon emission rate per unit solid angle for a line of two level atoms polarized in the $\boldsymbol{\hat{x}}$ direction:

\begin{equation}\label{eq:direction1}
\frac{d\Gamma_{\gamma}}{d\Omega} = \frac{3\Gamma}{8\pi}\Big\{1 - (\hat{\boldsymbol{k}}\cdot \hat{\boldsymbol{x}})^{2}\Big\}\sum_{n,m}e^{i\vec{\boldsymbol{k}}\cdot (\vec{\boldsymbol{r}}_{m}-\vec{\boldsymbol{r}}_{n})}a_{n}a_{m}^{*}
\end{equation}
where $\Gamma_{\gamma}$ is the photon scattering rate. For a periodic array of atoms parallel to the $\hat{\boldsymbol{z}}$ direction this becomes:
\begin{equation}\label{eq:direction2}
\frac{d\Gamma_{\gamma}}{d\Omega} = \frac{3\Gamma}{8\pi}\Big\{1 - \sin^{2}\theta\cos^{2}\phi\Big\}|a_{0}|^{2}\sum_{n,m} e^{i(m-n)\delta r(k\cos\theta - q)},
\end{equation}
where $a_{0}$ is a magnitude determined by the detuning and the Rabi frequency, while $q$ is the quasi-momentum of the eigenmode. All of the phases in Eq.~(\ref{eq:direction2}) will add coherently, resulting in a Bragg diffraction peak when:
\begin{equation}\label{eq:cond}
\delta r(k\cos\theta - q) = 0, \pm2\pi, \pm4\pi, ...,
\end{equation}
which means there will be a peak at the angle:
\begin{equation}\label{eq:angle}
\theta = \arccos\Big\{\lambda\Big(\frac{m}{\delta r} + \frac{q}{2\pi}\Big)\Big\}; m=0, \pm 1, \pm 2...
\end{equation}
If every atom has the same phase ($q=0$), then for values of $n\lambda < \delta r < (n+1)\lambda$ there will be $2n+1$ values of $\theta$ where Eq.~(\ref{eq:angle}) can be satisfied and photons may be scattered. Thus when $q=0$, the photon scattering of the array produces the well-known behavior of a diffraction grating. However, this is not the case when $q \ne 0$. For example when $0 < \delta r < \lambda/2$ and $q\delta r > k\delta r$, there are no solutions to Eq.~(\ref{eq:angle}). The result of this is that states with values of $q\delta r > k\delta r$ and $\delta r < \lambda/2$ \textit{do not decay}. This is seen in Fig. \ref{fig:band}(a), when $\Gamma_{q}$ jumps discontinuously to 0. The opposite effect happens when $\lambda/2 < \delta r < \lambda$. Here for small values of $q$, there is only one angle where the array can emit radiation coherently. However, when the value of $q$ is increased to the point where $\delta r(k + |q|) > 2\pi$ there is suddenly another value of $\theta$ corresponding to a diffraction peak, resulting in a discontinuous increase in the value of $\Gamma_{q}$ (see Fig. \ref{fig:band}). This pattern continues for larger values of $\delta r$ as well. If $\delta r = m\lambda + \eta$ ($m=0,1,2...$), where $\eta < \lambda/2$, then when $q$ is increased to the point where $q\delta r > k\eta$ there is one less diffraction peak where photons may escape. This makes the eigenmode's decay rate smaller. However if $\lambda/2 < \eta < \lambda$, when $|q\delta r| + k\eta > 2\pi$ there is one more allowed peak, making the decay rate of the eigenmode larger. For spatially disordered systems such as cold atom gases, this discontinuity does not occur. However when the atoms in a gas are excited to a timed-Dicke state, a similar peak still occurs in the forward direction \cite{scully2006} which also results in an increase in the decay rate \cite{sutherland2016, scully2007, scully2009}.

Another surprising feature of Fig. \ref{fig:band} is the divergence of the collective Lamb shift at the same values of $q$ where the discontinuities of $\Gamma_{q}$ occur. This happens because at this point the phase in the sum given by Eq.~(\ref{eq:band_eq}) becomes a multiple of $2\pi$, making the value of the imaginary part of $\epsilon_{q}$ dependent on a logarithmically diverging infinite sum over $\frac{1}{n\delta r}$. The decay rate does not diverge however. This is because when the phase in Eq.~(\ref{eq:band_eq}) becomes a multiple of $2\pi$ the real part of the $\propto 1/r$ term disappears, leaving only the convergent $1/r^{2}$ sum for the value of the decay rate.

\begin{figure}[h!]
	\includegraphics[width=0.45\textwidth]{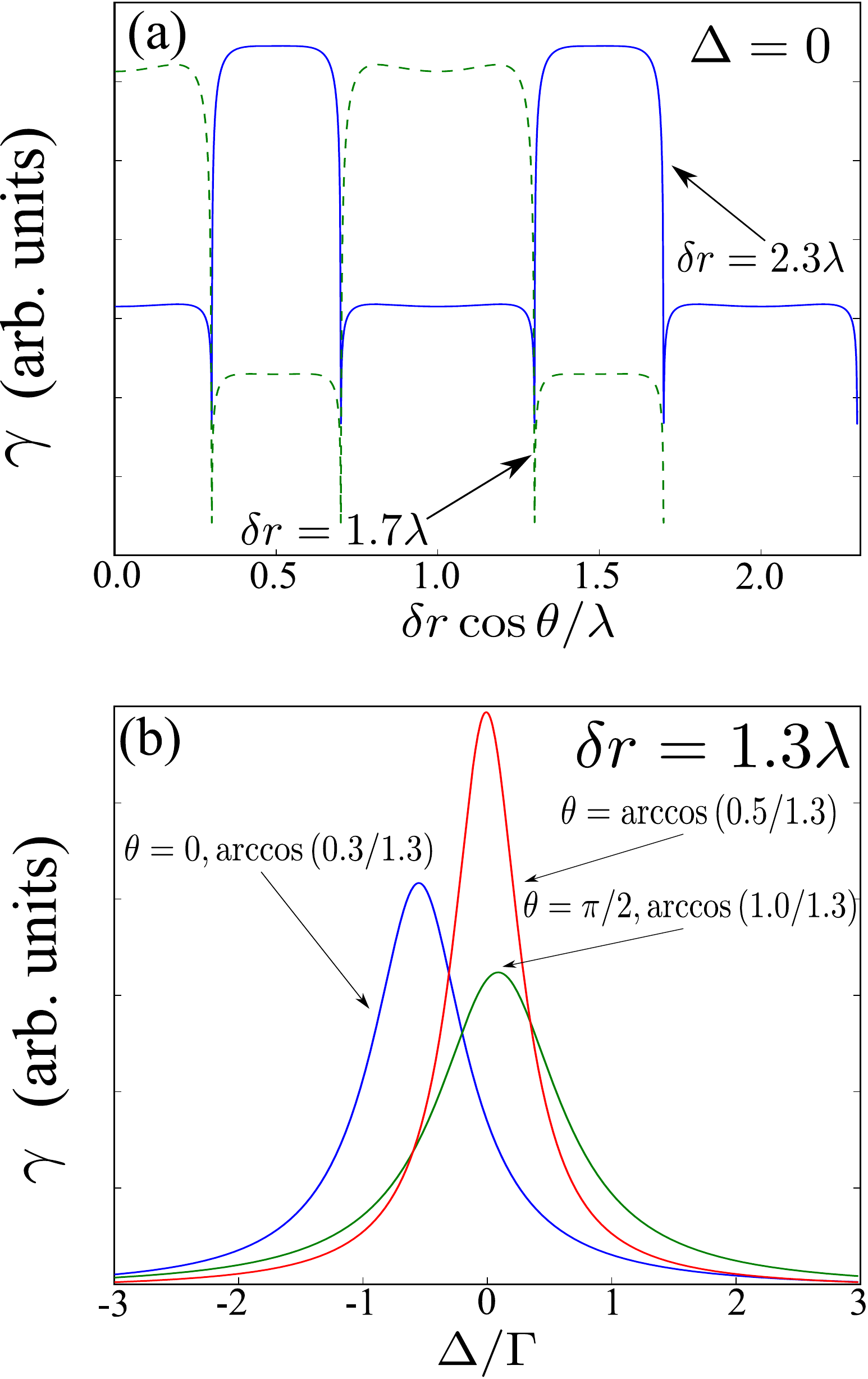}
	\caption{The dependance of the scattered photon emission on the laser angle $\theta$, which is the angle between an array of 1000 atoms and the driving laser. (a) The scattered radiation versus $\delta r\cos\theta/\lambda$ for a laser with $\Delta =0$ and arrays with spacings of $1.7\lambda$ and $2.3\lambda$. Note that the large shifts in scattered radiation occur when the eigenmode that the laser drives crosses a band structure discontinuity. (b) The photon scattering rate versus detuning of an array where $\delta r = 1.3\lambda$, for three different laser angles ($\theta$): \{0, $\arccos{(0.3/1.3)}$\}, \{$\pi/2$, $1.0/1.3$\}, and $\arccos{(0.5/1.3)}$. The scattering rates for all three angles fit to a Lorentzian lineshape well, with decay rates and lineshifts that correspond to those given by their $N\rightarrow \infty$ limit bandstructure calculations, seen in Fig. \ref{fig:band}.}
	\label{fig:angle}
\end{figure}

The band structure of an array has a strong effect on the scattered radiation versus the angle of the driving laser. This is because if one changes the angle of the laser, to first order they are also changing the relative phase correlations of the radiating atoms with respect to one other. For an array aligned along $\boldsymbol{\hat{z}}$, the phase of the laser that the $n^{th}$ atom sees is $e^{ikn\delta r\cos\theta}$. When $\theta$ is changed, the effective value of $q\delta r$ of the eigenmode that the laser is driving also changes. For example if the laser is perpendicular to the atomic array, the eigenmode corresponding to $q \delta r = 0$ is being driven. If the laser is situated at some arbitrary angle $\theta$, the value of $q\delta r$ of the eigenmode being driven is equal to $k\delta r \cos\theta$. Figure \ref{fig:angle}(a) shows how the photon scattering rate changes with respect to $\theta$. Here large changes in the scattering rate occur when the laser changes from driving a subradiant (superradiant) band to a superradiant (subradiant) band. Figure \ref{fig:angle}(a) also shows that the scattering rate dramatically drops at the point where this change happens, due to the large collective Lamb shift for this value of $q\delta r$. It should be noted that increases in $q\delta r$ of $2\pi$ correspond to the same phase correlations. For example, Fig. \ref{fig:angle}(b) shows that for an array with spacings $\delta r = 1.3\lambda$, $\theta = 0$ and $\theta = \arccos{(1.0/1.3)}$ give the exact same lineshape.

For values of $\delta r > \lambda/2$, the photon scattering lineshapes versus laser detuning fit a Lorentzian profile almost perfectly, with decay rates and collective Lamb shifts corresponding to their band structure values. As seen in Fig. \ref{fig:angle}(b) for an array of 1000 atoms and $\delta r = 1.3\lambda$, when $\theta = 0$ or $\theta = \arccos{(0.3/1.3)}$, $\Gamma_{q} \simeq 0.94\Gamma$ and $\Delta_{q} \simeq -0.54\Gamma$. The large shift in resonant energy occurs because of the logarithmically diverging collective Lamb shift discussed above, while $\Gamma_{q} \simeq 0.94\Gamma$ because the value of $q\delta r$ occurring at the discontinuous jump between subradiant and superradiant modes is driven, giving a value in the middle of the two. When $\theta = \pi/2$ or $\theta = \arccos{(1.0/1.3)}$, the superradiant band is driven, resulting in a broadened lineshape with $\Gamma_{q}\simeq 1.2$ and $\Delta_{q}\simeq 0.1\Gamma$, while when $\theta = \arccos{(0.5/1.3)}$ the subradiant band is driven, giving $\Gamma_{q}\simeq 0.66\Gamma$ and $\Delta_{q} = 0.01\Gamma$. This indicates that one may easily access subradiant eigenmodes by changing the angle of the laser.

\section{Conclusion} \label{conclusion}
The effects caused by a coherently radiating atomic array being driven by a laser have been studied. It was shown numerically that the excitation distribution for a given array is highly dependent on whether the driving laser is red or blue detuned, where the probability of excitation either grows or diminishes along $\boldsymbol{\hat{k}}$ respectively. It was also shown that the probability distribution and photon scattering become symmetric about the center of the array for spacings of $j\lambda/2$, where $j$ is an integer. It was then shown analytically that at large distances, the probability of excitation either grows or diminishes logarithmically no matter how large the value of $\delta r$ is. These results break down when $\delta r < \lambda/2$ due to the presence of a collection of extremely subradiant states ($\Gamma_{n}\sim 0$).

In order to interpret the eigenmodes of the system, a bandstructure calculation for an infinitely long array of atoms was conducted. These calculations showed the eigenmodes have both a collective Lamb shift that diverges logarithmically, as well as a discontinuous decay rate when plotted along the Brillouin zone. The sudden jump in decay rate can be understood by the appearance and disappearance of Bragg diffraction peaks of the scattered radiation. Finally, it was shown that these divergences and discontinuities in the bandstructure of an array may be exploited in order to manipulate the photon scattering rate by changing the angle of the driving laser, allowing one to manipulate the collective Lamb shift as well as access subradiant eigenmodes.

It has been suggested that coherently radiating systems should be thought of in terms of Bragg scattering \cite{porras2008, kaiser2016, hilliard2008}, where the radiators have certain spatially dependent phase correlations. Here, this picture is necessary. This is because the symmetries in an atomic array cause number of diffraction peaks to change. Since photons are mainly emitted from these diffraction peaks \cite{porras2008}, the photon scattering rate changes drastically with their appearance and disappearance. In systems such as cold atom gases, where the spatial distribution of atoms is highly disordered, the spatially dependent phases caused by the driving laser still cause a diffraction peak in the forward direction. This phenomenon is, of course, the well known Coherent Forward Scattering \cite{scully2007,rouabah2014}. A large part of understanding the nature of radiators, in the low excitation regime essentially consists of determining the spatially dependent phase relationship between atoms followed by determining their resulting scattered emission. Recently interesting new physics has resulted from approximating this relationship as the one caused by the initial driving laser, i.e. the timed Dicke state \cite{scully2007,scully2009,bromley2016,sutherland2016}. Understanding the nature of how these phase relationships develop, and how they may be manipulated in order to explore new physical systems will surely provide novel insights into the relationship between light and matter in the future. 

The authors would like to thank Prof. C.-L. Hung for enlightening conversations, as well as Jes\'{u}s P\'{e}rez-R\'{i}os for his valuable insights into graphic design. This material is based upon work supported by the
National Science Foundation under Grant No. 1404419-PHY.

\bibliography{bibtex.bib}
\end{document}